\documentclass[peerreview,draftcls,onecolumn,12pt,letterpaper]{IEEEtran}

\ifCLASSINFOpdf
\else
\fi

\usepackage{graphicx}

\usepackage{url}
\usepackage{subfigure}        
\usepackage[usenames]{color}            
\usepackage{amsfonts}%
\usepackage{amssymb}%
\usepackage{mathrsfs}
\usepackage{amsmath}%
\usepackage{cite}%

\usepackage[latin1]{inputenc}

\newcommand{\EE}{\mathbf{E}}
\newcommand{\ud}{\mathrm{d}}
\begin{document}
\title{Zero-Delay Joint Source-Channel Coding for a Bivariate Gaussian on a Gaussian MAC}

\author{Pål~Anders~Floor, Anna~N.~Kim, Niklas~Wernersson, Tor~A.~Ramstad, Mikael~Skoglund and Ilangko~Balasingham,
\thanks{P. A. Floor  and I. Balasingham is at the Interventional Center,
Oslo University Hospital and Institute of Clinical Medicine, University of Oslo,
  Oslo, Norway (e-mail: andflo@rr-research.no). A. N. Kim,  T. A.
Ramstad  and I. Balasingham is with the Department of Electronics and
Telecommunication, Norwegian University of Science and Technology
(NTNU), Trondheim,
Norway (e-mail: annak@iet.ntnu.no).}
\thanks{Niklas Wernersson is with Ericsson Research, Stockholm, Sweden.
Mikael Skoglund is with the School of Electrical Engineering and the ACCESS Linnaeus Center,
at the Royal Institute of Technology (KTH), Stockholm, Sweden}
\thanks{This work was supported by the Research Council of Norway (NFR), under the projects MELODY nr. 187857/S10 and CROPS2 nr. 181530/S10.
The work of M. Skoglund was funded in part by the Swedish Research Council and VINNOVA.}}
\maketitle

\begin{abstract}
In this paper, delay-free, low complexity, joint
source-channel coding (JSCC) for transmission of two correlated
Gaussian memoryless sources over a Gaussian Multiple Access Channel
(GMAC) is  considered. The main contributions of the paper are two
distributed JSCC schemes: one discrete scheme based on nested scalar quantization, and one hybrid
discrete-analog scheme based on a scalar quantizer and a linear
continuous mapping.
The proposed schemes show promising performance
which improve with increasing correlation and are robust
against variations in noise level. Both schemes exhibit a constant gap to the performance upper bound when the channel signal-to-noise ratio gets large.
\end{abstract}
%
\section{Introduction}
In a point-to-point communication system, for a memoryless Gaussian
source-channel pair of equal bandwidth, it is well known that a simple encoder that scales its incoming signal to satisfy the channel power constraint and
\emph{minimum mean square error} (MMSE) decoding at the receiver achieves the information theoretical bound \emph{optimal performance theoretically
attainable} (OPTA)~\cite{Goblick65}. This linear approach, which is often referred to as \emph{uncoded transmission} in the literature, constitutes a very simple joint
source-channel coding (JSCC) scheme due to its low complexity and
zero coding delay. Separate source and channel coding (SSCC), as
summarized by the \emph{separation theorem}~\cite{Shannon48mathematical}, also achieves
OPTA, but requires infinite complexity and delay.

In this paper we investigate a multipoint-to-point problem, where
two memoryless and inter-correlated Gaussian sources are
transmitted over a memoryless Gaussian multiple access channel (GMAC).
There are mainly two cases to consider for such a network:
1) Recovery of the common information shared by the two sources. 2) Recovery
of each individual source.

Case 1) was studied in e.g.,~\cite{Gastpar08},\cite{Gastpar07},\cite{Behroozi09}.
Distortion lower bounds was derived and for the case of equal (source) variance, it was shown that
these bounds are achieved by uncoded transmission when the transmit power of all encoders are equal. It was also
shown that any \textit{distributed} SSCC scheme is sub-optimal, except when the sources are uncorrelated.

Case 2) was recently studied
in~\cite{Lapidoth10}. The distortion lower bound was derived
by allowing full collaboration between the encoders and thus
converting the multi-point-to-point into a point-to-point
communication problem. The best possible performance can then be
determined by bounding the rate-distortion region of the bivariate
Gaussian~\cite[Theorem III.1]{Lapidoth10} by the GMAC's \emph{sum
rate}~\cite[Theorem IV.1]{Lapidoth10}. Closed form solutions were
given for the \emph{symmetric case}~\cite[Corollary IV.1]{Lapidoth10}, i.e. when the
average distortion in the reconstruction and the transmit power are equal for both sources.
It was shown that uncoded
transmission achieves the distortion lower bound up to a certain
channel signal-to-noise ratio (SNR), depending on the correlation
between the two sources. In order to get close to the distortion
lower bound in general, the authors further proposed a nonlinear hybrid
scheme that superimposes a rate optimal (infinite dimensional) vector quantizer (VQ) and uncoded
transmission. The hybrid scheme was shown to be optimal at high and low channel
SNR, while a small gap remains for other SNRs. The
authors also examined distributed SSCC. Just as in case 1), SSCC is sub-optimal
except when the sources are uncorrelated. Contrary to case 1),
optimality of uncoded transmission for case 2) is
restricted and infinite complexity and delay JSCC is required to close in on the bounds in general.

This result prompts an important question: What happens if we
impose a strict complexity and delay constraint for case 2)? The main
objective of this paper is to provide answers to this question. In
particular, we want to find well performing, simple and
implementable schemes with zero coding delay, just as that offered by uncoded transmission. That is, simple distributed nonlinear mappings that offer better performance than uncoded transmission outside the SNR domain where uncoded transmission is optimal.

It is important to keep in mind that
when assessing performance of a JSCC scheme designed under a
zero-delay constraint, one must expect a significant backoff from the
distortion lower bound derived in \cite{Lapidoth10}, as it is based
on infinite block length. We therefore briefly address known zero-delay cooperative\footnote{By cooperation we mean that both
source symbols are available at both encoders without any additional use of resources.} encoding schemes to provide
 indications on where the bound may lie for any zero-delay distributed scheme.

The rest of the paper is organized as follows: In
Section~\ref{sec:prob_def}, the problem formulation
and relevant bounds are given. Zero-delay mappings for cooperative encoding are also introduced. In
Section~\ref{sec:distributed_delayfree}, we present two distributed
schemes: A discrete (digital) mapping based
on \emph{Nested Quantization}~\cite{Servetto07}, and a hybrid discrete-analog scheme using a scalar quantizer and a limiter followed by a linear
coder. Both schemes are optimized. Extensions to more general
cases are discussed and we show that the suggested schemes exhibits a constant gap to the bound when SNR$\rightarrow\infty$. In Section~\ref{sec:perf_and_sim}, the
proposed schemes are simulated and compared to the performance
bounds and other relevant schemes under both average- and equal transmit
power constraints. We summarize the results in the paper in
Section~\ref{sec:summary} and give some future research
directions.
\section{Problem statement and upper bounds}\label{sec:prob_def}
\begin{figure}[h!]
\begin{center}
\hspace{0.5cm}
 \includegraphics[width=0.9\columnwidth]{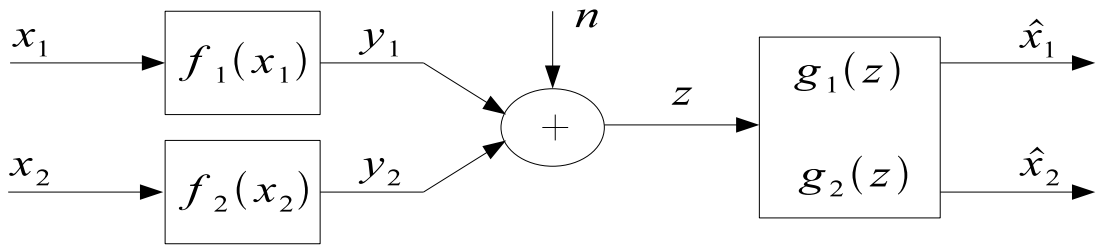}
\end{center}
  \caption{Two correlated Gaussian memoryless sources
  transmitted on a GMAC with both sources reconstructed at the receiver.}\label{fig:system}
\end{figure}

The communication system under consideration is shown in Fig.~\ref{fig:system}.
\subsection{Problem statement}\label{ssec:prob_def_a}
The sources $x_1$ and $x_2$ are assumed to be zero mean Gaussian random variables $x_m=v+w_m \sim\mathcal{N}(0,\sigma_{x_m}^2)$, $\hspace{0.5cm}m=1,2,$
 where $v\sim\mathcal{N}(0,\sigma_v^2)$ is the common information for
both sources and $w_m\sim\mathcal{N}(0,\sigma_{w_m}^2)$ is unique to
each source, where $v$ and $w_m$ are independent. Further we assume
that $\sigma_{w_1}^2=\sigma_{w_2}^2$, implying that the variances
 $\sigma_{x_1}^2=\sigma_{x_2}^2=\sigma_x^2$. The $2\times 2$ covariance matrix $C_x$ will then have $\sigma_x^2$ on the diagonal and $\sigma_x^2\rho_x$ on the off-diagonal elements
with eigenvalues $\lambda_1=\sigma_x^2 (1 + \rho_x)$ and $\lambda_2=\sigma_x^2 (1 - \rho_x)$, where $\rho_x= \EE[x_1x_2]/\sigma_x^2=\sigma_v^2/\sigma_x^2$ denote the
correlation between $x_1$ and $x_2$. The joint probability density
function (pdf) is given by
\begin{equation}\label{e:joint_pdf}
p_{\mathbf{x}}(\mathbf{x})=p_{\mathbf{x}}(x_1,x_2)=\frac{1}{2\pi\sigma_x^2\sqrt{1-\rho_x^2}} e^{-\frac{1}{2}\mathbf{x}^T C_x^{-1} \mathbf{x}}
\end{equation}
with equal marginals $p_x(x_m)$.

We denote the zero-delay encoding functions by $f_m(x_m), m=1,2$. The average transmit power from encoder $m$ is then $P_m =
\EE[|f_m(x_m)|^2], m = 1,2$. The encoder outputs are transmitted on a memoryless MAC with additive Gaussian noise $n\sim\mathcal{N}(0,\sigma_n^2)$ with
pdf $p_n(n)$.

We assume ideal Nyquist sampling and an ideal Nyquist channel where
the sampling rate of each source is the same as the signalling rate
of the channel. We also assume ideal synchronization and timing
between all nodes.

The received signal
\begin{equation}\label{e:ch_sign}
z=f_1(x_1)+f_2(x_2)+n
\end{equation}
is passed through the decoding functions $g_1,g_2$ to produce
an estimate of each individual source $\hat{x}_1,\hat{x}_2$.
We use the mean-squared-error distortion
criterion, and define the \textit{average} end-to-end distortion as:
\begin{equation}\label{e:tot_dist}
D = \frac{1}{2}(D_1+D_2)=\frac{1}{2}\big(\EE\{|x_1-\hat{x}_1|^2\}+\EE\{|x_2-\hat{x}_2|^2\}\big).
\end{equation}
Our goal is to design the mapping functions $(f_m,g_m),
m=\{1,2\}$, under a given power constraint, so that $D$ is
minimized.
\subsection{Performance upper bound}\label{ssec:prob_def_b}
For the above defined communication system, we consider both an average
and equal transmit power constraint. The reason for the former is
that the distributed schemes we propose in
Section~\ref{sec:distributed_delayfree} have asymmetric encoders resulting in $P_1\geq P_2$. It is
therefore convenient to optimize our schemes under an average
transmit power constraint $P$, where $P_1+P_2 = 2P$. If an equal transmit power constraint $P_1=P_2=P$ is imposed
a loss is expected (shown in Section~\ref{ssec:sim_equal_power}).

The performance upper bound for the \emph{symmetric case} $D_1=D_2=D$ and $P_1=P_2=P$, expressed in terms of the signal-to-distortion ratio (SDR)
is
\begin{align}\label{e:OPTA_equalP}
SDR=\frac{\sigma_x^2}{D}=
\begin{cases}
\big(\frac{P(1-\rho_x^2)+\sigma_n^2}{2P(1+\rho_x)+\sigma_n^2}\big)^{-1}, & \frac{P}{\sigma_n^2} \in\big(0,\frac{\rho_x}{1-\rho_x^2}\big],\vspace{0.3cm}\\
\big(\frac{\sigma_n^2(1-\rho_x^2)}{2P(1+\rho_x)+\sigma_n^2}\big)^{-\frac{1}{2}}, & \frac{P}{\sigma_n^2}> \frac{\rho_x}{1-\rho_x^2},\end{cases}
\end{align}
where $D$ is the distortion lower bound from~\cite[Corollary IV.1]{Lapidoth10}. Although this bound was derived for $P_1=P_2=P$ it is also a bound for an average transmit power constraint by simply substituting $P=(P_1+P_2)/2$.
The term ``channel SNR" refers to $P/\sigma_n^2$ in the rest of the paper. Uncoded transmission achieves the bound given by the upper
equation in~(\ref{e:OPTA_equalP})~\cite[Corollary IV.3]{Lapidoth10}, and corresponds to the SNR where only the common
information $v$ can be recovered at the decoder.
\subsection{Delay-free JSCC for cooperative encoders}\label{sec:nonlin_cooperative}
Since collaboration makes it possible to construct a larger set of encoding operations, including all distributed strategies, the performance of distributed
coding schemes are upper-bounded by those that allow collaboration when properly optimized. The performance of the
proposed distributed JSCC scheme relative to the performance
upper-bound described in \cite{Lapidoth10} is one such example.

In the case of zero delay, the corresponding optimal collaborative
encoding operation is the optimal mapping $\mathbb{R}^2\rightarrow \mathbb{R}$
from source to channel space, which minimizes $D$ at a given power
constraint. Finding the optimal structure of such a mapping is a problem yet to be solved. We can, however, get an
idea of how collaborative encoders may perform from known schemes that
perform close to the bounds. Examples on schemes with excellent performance are
\emph{Shannon-Kotel'nikov mappings} (S-K mappings)~\cite{hekland_floor_ramstad_T_comm},~\cite{Akyol_rose_ramstad_itw10} and  \emph{Power Constrained Channel
Optimized Vector
Quantizers} (PCCOVQ)~\cite{fulds97a}. When the number of centroids in the PCCOVQ is large they are very similar to S-K mappings, we therefore refer to both of these as S-K mappings in the following.
S-K mappings have previously been optimized for
memoryless Gaussian sources and channels when  $M$ source symbols are
transmitted on $N$ channel uses~\cite{hekland_floor_ramstad_T_comm,fulds97a,FloorThesis,Hu_garcia_lamarca09}.

For the problem at hand, if we treat the collaborative encoders as one, and the two sources as two components of
a Gaussian vector source, we can apply S-K mappings with $M=2$ and $N=1$ directly.
This collaborative zero-delay scheme can not be applied to the
distributed case as their operation relies on knowing both
source symbols simultaneously at each encoder~\cite{hekland_floor_ramstad_T_comm,fulds97a}. Nevertheless, we can use the
collaborative schemes as benchmarks and see how much loss one may
expect with distributed encoding. This will be illustrated in Section \ref{sec:perf_and_sim}.
%
%
%
\section{Distributed zero-delay schemes}\label{sec:distributed_delayfree}
Since we want to split the two interfering sources at the
receiver, we have to construct our JSCC schemes accordingly. A discrete (digital) approach that achieves this purpose is \emph{Nested Quantization}
(NQ) introduced in~\cite{Servetto07} with sequential decoding at the receiver.

\subsection{Nested Quantization}\label{ssec:nest_q}
\begin{figure}[h!]
    \begin{center}
          \includegraphics[width=1\columnwidth]{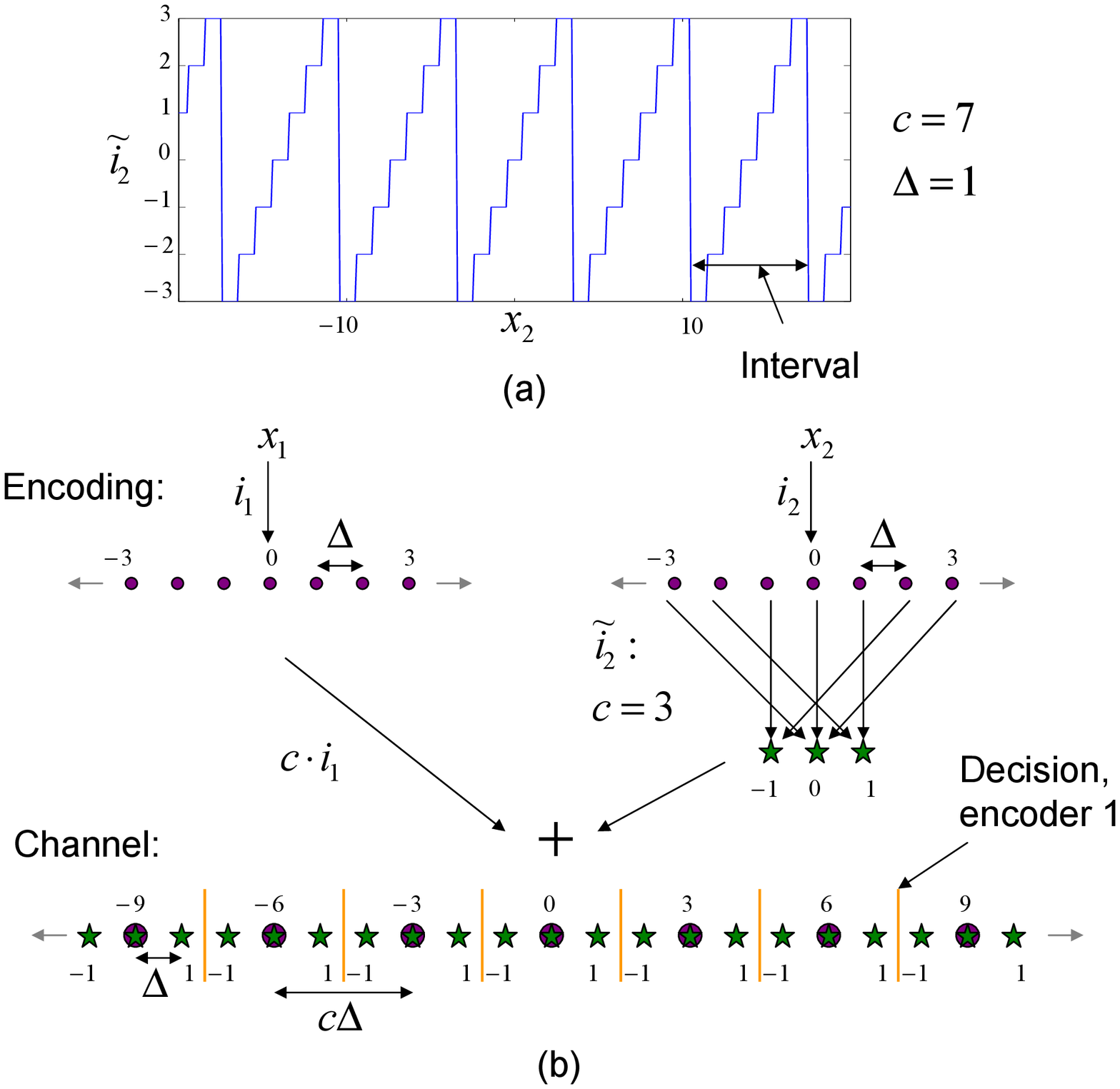}
    \end{center}
    \caption{NQ encoding process. (a) The nested quantizer $\tilde{i}_2$ for $c=7$ and $\Delta=1$. The whole real line, representing $x_2$, is mapped to a fixed number of centroids (here 7 values) inside a finite interval. (b) How encoding is performed when $c=3$. The gray arrows indicate that the structure continues.}
    \label{fig:nq_enc_2}
\end{figure}

Our NQ scheme consists of two uniform scalar quantizers, one for
each encoder.
Without loss of generality we choose encoder 2 to be the \emph{nested quantizer}. Fig.~\ref{fig:nq_enc_2} shows the NQ encoding process. 
The encoding functions $f_1(x_1)$ and $f_2(x_2)$ are\footnote{ The following equations describe
midthread quantizers with odd $c$. Similar equations
can be derived for midrise quantizers.}
\begin{equation}\label{e:enc_NQ}
y_1=f_1(x_1)=a\cdot c\cdot i_1(x_1),\ \ \ y_2=f_2(x_2)=a\cdot\tilde{i}_2(x_2).
\end{equation}
$i_m(x_m)$ denotes the quantization process that returns an index in
$\mathbb{Z}$:
\begin{equation}\label{e:quantizer}
i_m(x_m)=\bigg\lfloor\frac{x_m}{\Delta}\bigg\rceil, \ \ m=1,2,
\end{equation}
where $\lfloor\cdot \rceil$ denotes rounding to the nearest integer
and $\Delta$ denote the quantization step.
Further, for encoder 2, the nested quantizer is invoked by
\begin{equation}\label{e:nest_map}
\tilde{i}_2(x_2)= i_2(x_2)-c\bigg\lfloor\frac{i_2(x_2)}{c}\bigg\rceil, \hspace{1cm} c\in\mathbb{N},
\end{equation}
as depicted in Fig.~\ref{fig:nq_enc_2}(a). I.e., $\tilde{i}_2$ is the mapping
\begin{equation}\label{e:enc2_nq_map}
\tilde{i}_2: \mathbb{R}\overset{i_2}{\rightarrow} \mathbb{Z} \rightarrow \left\{-\frac{c-1}{2}, -\frac{c-3}{2}, \cdots
\frac{c-1}{2}\right\}.
\end{equation}
That is the whole real line is mapped into a fixed bounded interval of discrete values.
In order to identify each encoder output after they have been summed over the GMAC,
$i_1$ must be scaled by $c$. This is illustrated in Fig.~\ref{fig:nq_enc_2}(b) after the sum. Now both sources can be reconstructed at the receiver through sequential decoding.
The parameter $a$ is set to satisfy the power constraint.
The average transmit power from each encoder is
\begin{equation}\label{e:power_nq}
  P_1 =  c^2 a^2\sum_{i_1 = -\infty}^{\infty} Pr(i_1)i_1^2, \hspace{1cm}
  P_2 = a^2 \sum_{\tilde{i}_2 = -(c-1)/2}^{(c-1)/2} Pr(\tilde{i}_2)\tilde{i}_2^2
\end{equation}
where $Pr(\cdot)$ denote the probability for the event inside the parentheses.

At the decoder
we first recover the indices from encoder 1, $j_1(z)$, using
the maximum likelihood (ML) estimate (based on the decision values seen in Fig.~\ref{fig:nq_enc_2})
\begin{equation}\label{e:decode_one}
\begin{split}
j_1(z)&= \arg\min_{j\in \mathbb{Z}}\|f_1(j\Delta)-z\|^2.
\end{split}
\end{equation}
$\tilde{j}_2(z)$ from encoder 2 is then detected after
subtracting $f_1(j_1(z)\Delta)$ from the channel output
\begin{equation}\label{e:decode_two}
\begin{split}
\tilde{j}_2(z)&=\arg \min_{j\in\mathbb{Z}:|j|\leq\frac{c-1}{2}}\|z-f_1(j_1(z)\Delta)-f_2(j\Delta)\|^2.
\end{split}
\end{equation}
As seen from Fig.~\ref{fig:nq_enc_2}(a), $\tilde{j}_2(z)$ corresponds to an infinite number of estimates of $i_2$. It is therefore not
possible to invert the mapping (\ref{e:enc2_nq_map}) from
$\tilde{j}_2$ to $j_2$ based on $\tilde{j}_2(z)$ alone. But the
decoder also has access to $j_1(z)$, which due to the correlation
contains information about $x_2$, and hence also $j_2(z)$.
To determine the most likely \emph{interval} (see Fig.~\ref{fig:nq_enc_2}(a)) that $j_2$ belongs to, we consider the maximization of~(\ref{e:joint_pdf}).
Given $x_1=j_1(z)\Delta$ and that $\arg \max_{\mathbf{x}} p_\mathbf{x}(\mathbf{x})= \arg \min_{\mathbf{x}} \left(\mathbf{x}^T C_x^{-1} \mathbf{x}\right)$
the most likely value of $j_2$ can be approximated by
\begin{align}
    \label{eq:choose_j}
    j_2(z) = \arg \min_{j_2} \left(
\left[\begin{array}{c}
  j_1(z)\Delta \\
  j_2 \Delta \\
\end{array}\right]^T C_x^{-1} \left[\begin{array}{c}
  j_1(z)\Delta \\
  j_2 \Delta \\
\end{array}\right]\right),
\end{align}
where $j_2 \in \{\tilde{j}_2(z) + k c, k\in
\mathbb{Z}\}$. Solving~(\ref{eq:choose_j}) with respect to $k$ we get
\begin{equation}\label{e:optimal_cell_choice}
k=\max\bigg\{\bigg\lfloor\frac{\rho_x j_1(z)-\tilde{j}_2(z)}{c}\bigg\rceil,0\bigg\}
\end{equation}
By choosing the constant $c$
appropriately with respect to $\rho_x$, the decoder will
with high probability be able to invert the mapping $i_2
\rightarrow\tilde{i}_2$ correctly.
The larger $\rho_x$ is, the more accurately $j_1(z)$ will help in telling which interval $i_2$ is in, implying that $c$ can be made smaller the larger $\rho_x$ is.

In order to minimize the MSE, we further calculate
\begin{equation}\label{e:MMSE_dec}
\hat{x}_m=\EE\{x_m|j_1(z),j_2(z)\},\hspace{0.5cm}m=1,2.
\end{equation}

To design the optimal NQ, we need to determine the $\Delta$, $c$ and
$a$ that minimize $D$ under a given power constraint.
Since the two encoders are asymmetric, $P_1\geq P_2$, as seen from~(\ref{e:power_nq}).
We must solve the following optimization problem
\begin{equation}\label{e:opt_problem_nq}
\min_{\Delta,c,a : P_1+P_2\leq 2P} D,
\end{equation}
where $D$ is the average end-to-end distortion and $P$ is
the average transmit power.

\textbf{Distortion calculation:}
In order to simplify notation when deriving the distortion, we introduce the following auxiliary variables
\begin{equation}\label{e:aux_var}
\bar{x}_m=\EE\{x_m|i_1,i_2\},\hspace{0.5cm} \tilde{x}_m=\EE\{\hat{x}_m|i_1,\tilde{i}_2\},\hspace{0.5cm}m=1,2.
\end{equation}
$\bar{x}_m$ is the quantized source, $\tilde{x}_m$ is the quantized
source after the map $\tilde{i}_2$ and $\hat{x}_m$ is as defined
in~(\ref{e:MMSE_dec}). The Appendix shows that the per source distortion $D_m$ can be split into three terms:
\begin{equation}\label{e:Dtot_discrete}
\begin{split}
D_m &= \EE\{|x_m-\hat{x}_m|^2\}=\EE\{|x_m-\bar{x}_m|^2\}+\EE\{|\bar{x}_m-\tilde{x}_m|^2\}+\EE\{|\tilde{x}_m-\hat{x}_m|^2\}\\
&=\bar{\varepsilon}_{q,m}+\bar{\varepsilon}_{c,m}+\bar{\varepsilon}_{n,m}.
\end{split}
\end{equation}
$\bar{\varepsilon}_{q,m}$ is the quantization distortion given by
\begin{equation}\label{e:nq_qdist}
\begin{split}
 \bar{\varepsilon}_{q,m} & = \iint
    p_{\mathbf{x}}(x_1, x_2)(x_m - \bar{x}_l(i_1(x_1), i_2(x_2)))^2 \ud x_1 \ud
    x_2 \approx \frac{\Delta^2}{12}, 
\end{split}
\end{equation}
where the last approximations is valid for small $\Delta$.
$\bar{\varepsilon}_{c,m}$ represents the distortion from the inversion of ${i}_m\rightarrow \tilde{i}_m$, resulting when the wrong interval in Fig.~\ref{fig:nq_enc_2}(a) is detected at the decoder
\begin{equation}\label{e:nq_cdist}
\begin{split}
   \bar{\varepsilon}_{c,m} & = \iint
    p_{\mathbf{x}}(x_1, x_2) (\bar{x}_m(i_1(x_1), i_2(x_2))-\tilde{x}_m(i_1(x_1), \tilde{i}_2(x_2)))^2 \ud x_1 \ud x_2\\
    & = \sum_{i_1,i_2} Pr(i_1,i_2) (\bar{x}_m(i_1, i_2) -\tilde{x}_m(i_1, \tilde{i}_2(i_2)))^2,
\end{split}
\end{equation}
where $\tilde{i}_2(i_2)$ represents the mapping from $i_2$ to
$\tilde{i}_2$ in~(\ref{e:nest_map}).
$\bar{\varepsilon}_{n,m}$, the distortion due to channel noise, is
given by the following equation
\begin{equation}\label{e:nq_ndist}
\begin{split}
  \bar{\varepsilon}_{n,m}^2 &=\iiint p_{\mathbf{x}}(x_1,x_2) p(z | i_1(x_1), \tilde{i}_2(x_2)) 
  (\tilde{x}_m(i_1(x_1), \tilde{i}_2(x_2)) - \hat{x}_m (j_1(z),j_2(z)))^2 \ud z \ud x_1 \ud x_2\\
  &= \sum_{i_1, \tilde{i_2}} \sum_{j_1, j_2} Pr(i_1,\tilde{i}_2) Pr(j_1, j_2|i_1,\tilde{i}_2)
    (\tilde{x}_m(i_1, \tilde{i}_2) - \hat{x}_m(j_1,j_2))^2. 
\end{split}
\end{equation}
Since the NQ is discrete, the integrals in~(\ref{e:nq_qdist})-(\ref{e:nq_ndist}) can be written as sums, the
density function $p_\mathbf{x}(x_1,x_2)$ can be replaced by the
point probabilities $Pr(i_1,i_2)$ and
$p(z|i_1(x_1),\tilde{i}_2(x_2))$ will be fully determined by the
transition probabilities $Pr(j_1,j_2|i_1,\tilde{i}_2)$. The distortion in~(\ref{e:Dtot_discrete}) is a function of
all three parameters $\Delta$, $c$ and $a$: $i_1$ depends on
$\Delta$,  $i_2$ depends on $\Delta$ and $c$, and $j_1,j_2$ depends
on $\Delta,c$ and $a$, as seen from~(\ref{e:decode_one})
and~(\ref{e:decode_two}).

In the next section we propose a scheme where encoder 2
is continuous, i.e. a hybrid discrete analog scheme named \emph{Scalar Quantizer Linear Coder} (SQLC).
%
%
%
\subsection{Hybrid discrete-analog scheme: SQLC}\label{ssec:SQLC}
The motivation for introducing the SQLC in addition to the NQ are mainly given by the following two reasons: 1) The SQLC does not introduce quantization distortion at the encoder 2, implying that the SQLC should improve over the NQ (as confirmed in Section~\ref{sec:perf_and_sim}).
2) The SQLC is ideally simpler to implement since the encoder 2 basically consist of a limiter followed by scaling, whereas the NQ require two rounding operations.

Encoder 1 is now a midrise quantizer with an even number of levels, where the representation values are transmitted directly on the channel.
Encoder 2 is a limiter, denoted by $\ell_{\pm \kappa}[\cdot]$, that clips the amplitude of $x_2$ to $\pm \kappa$, $\kappa\in \mathbb{R}_+$,
followed by scaling with $\alpha$. That is,
\begin{equation}\label{e:sqlc_encoder}
f_1(x_1)= \Delta\bigg(\bigg\lfloor\frac{x_1}{\Delta}-\frac{1}{2}\bigg\rceil + \frac{1}{2}\bigg),\hspace{0.5cm} f_2(x_2)=\alpha \cdot(\ell_{\pm \kappa}[ x_2]),
\end{equation}
To simplify notation we denote centroid number $i$ of encoder 1 by $q_i$ in the following.

\begin{figure}[h!]
    \begin{center}
          \includegraphics[width=1\textwidth]{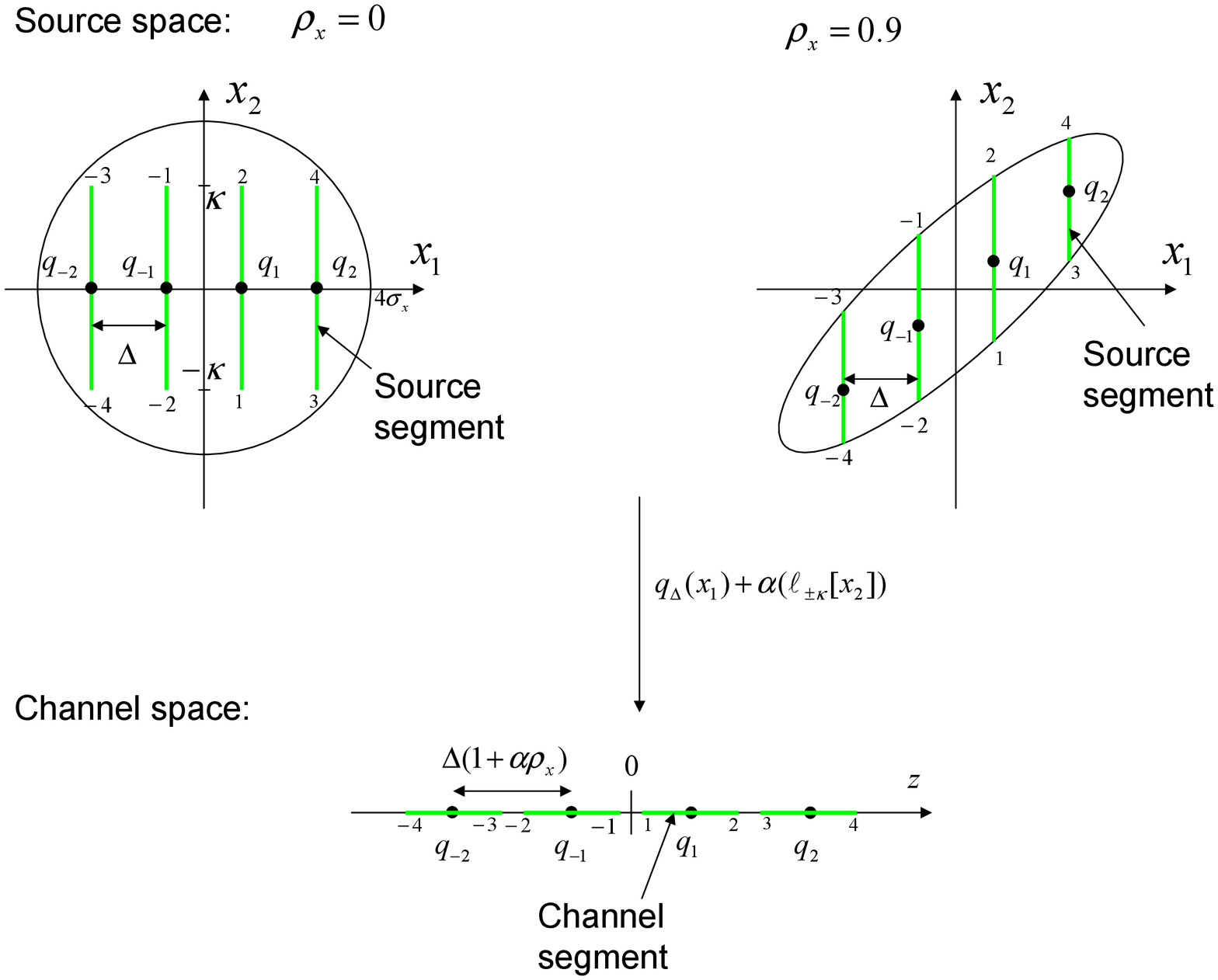}
    \end{center}
    \caption{SQLC concept for $\rho_x=0$ and $0.9$. The dots represent encoder 1 while the line segments represent encoder 2. The numbers show how the segments in the source and channel space are related. If $\alpha$, $\kappa$ and $\Delta$ are chosen appropriately in relation to each other, the channel segments will not overlap, and both sources can be decoded uniquely at the receiver.}
\label{fig:sqlc_config}
\end{figure}

$\alpha$ and $\kappa$ must be chosen small enough in relation to $\Delta$ so we achieve the geometrical configuration shown in Fig.~\ref{fig:sqlc_config} (depicted for $\rho_x=0$ and $0.9$), i.e. with non-intersecting channel segments. This will make it possible to uniquely decode both the quantized (depicted as dots) and continuous values (depicted as segments) from their sum.
Note that when $\rho_x$ increases,
the limitation to $\pm \kappa$ becomes more and more insignificant
since the joint pdf $p_\mathbf{x}(x_1,x_2)$ narrows along its minor axis, effectively ``limiting'' each source segment of the SQLC (the geometry of $p_\mathbf{x}(x_1,x_2)$ ``limits''
$x_2$ given $x_1$ when $\rho_x$ gets close to one), resulting in reduced distortion.

Sequential decoding is again applied. First source 1 is recovered as
\begin{equation}\label{e:decode_one_sqlc}
g_1(z)=\arg\min_{q_i}\|\mu(q_i)-z\|^2,
\end{equation}
where $\mu(q_i)$ takes into account that the
midpoint of each channel segment shown in Fig.~\ref{fig:sqlc_config}
changes with $\rho_x$. That is, given that centroid
$q_{i}$ was transmitted from encoder 1, the midpoint for the
relevant channel segment becomes
$q_{i}+E\{\alpha x_2 |q_{i}\} = q_{i}+ \alpha \rho_x q_{i},$ and so
\begin{equation}\label{e:ch_pdf_mean}
\mu = q_i (1+\alpha \rho_x).
\end{equation}
We used the relation $E\{x_2|x_1\}=\rho_x x_1$~\cite[p. 233]{papoulis02} for this calculation.
Source 2 is then decoded as
\begin{equation}\label{e:decode2_sqlc}
g_2(z)=\beta
(z-g_1(z)),
\end{equation}
where $\beta$ is an amplification factor. We further mimimize the MSE by computing
\begin{equation}\label{e:MMSE_dec_SQLC}
\hat{x}_m=\EE\{x_m|g_1(z),g_2(z)\},\hspace{0.5cm}m=1,2.
\end{equation}

We again formulate the optimization problem with an average transmit power constraint:
\begin{equation}\label{e:opt_problem}
\min_{\alpha,\beta,\Delta,\kappa : P_1+P_2\leq 2P} D.
\end{equation}

To calculate $D$ we could formulate similar integrals as for the NQ in Section~\ref{ssec:nest_q}. But since encoder 2 is now continuous, one will have to solve multiple integrals numerically. A less computationally consuming approach, which also gives additional insight into the SQLC's underlying principle, is to divide $D$ into five contributions: two from the encoding process and
three due to channel noise. From the encoding process, we get
quantization distortion $\bar{\varepsilon}_{q}^2$ for source 1 and
clipping distortion $\bar{\varepsilon}_{\kappa}^2$ for source 2. The
effect of channel noise on source 1 is only present when the
centroids of encoder 1 are mis-detected, and we named it \emph{channel
distortion} $\bar{\varepsilon}_{Ch1}^2$. Inspired by~\cite[pp.62-98]{kotelnikov59}, we further divide the effect of channel
noise on $x_2$ into additive noise, which name
channel distortion $\bar{\varepsilon}_{Ch2}^2$, and \emph{anomalous
distortion} $\bar{\varepsilon}_{an}^2$, resulting from a
\emph{threshold effect}~\cite{shann49,Merhav_N}, that leads to large
decoding errors. Here, threshold effect results when
the wrong centroid for encoder 1 is detected since we will jump from one
channel segment to another (see Fig.~\ref{fig:sqlc_config}).

\begin{figure}[h!]
    \begin{center}
            \vspace{1cm}
            \includegraphics[width=0.8\columnwidth]{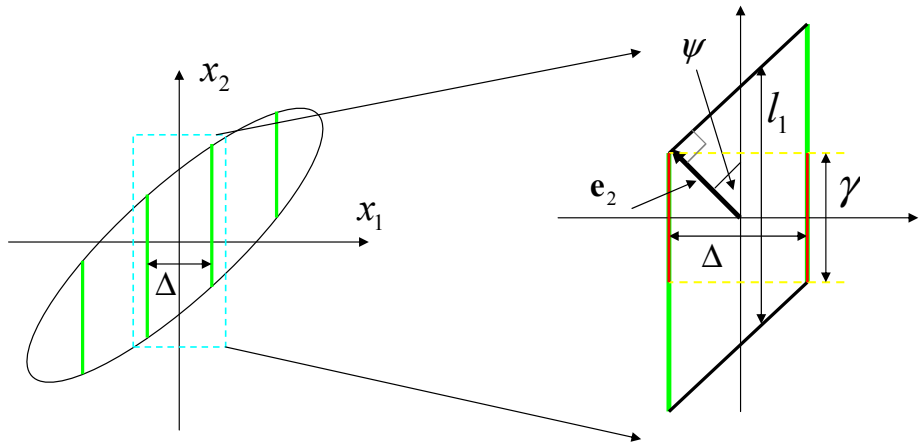}
       \end{center}
    \caption{SQLC in source space when $\rho_x= 0.9$. The enlarged portion shows how to approximately calculate anomalous errors for $x_2$ when $\rho_x$ is \emph{high} ($>\approx 0.7$). $l_1$ denotes the length of the portion of the $x_2$ axis that contains the significant probability mass given $x_1$, and $\mathbf{e}_2$ denote the minor axis of the ellipse shown (source space). }\label{fig:SQLC_corr_concept}
\end{figure}

In order to calculate the effect of channel noise on the distortion, we first need the channel
output pdf. We refer to Fig.~\ref{fig:SQLC_corr_concept} and~\ref{fig:sqlc_config} in order to explain the derivation of the pdf.

\subsubsection{Channel output pdf}\label{ssec:derive_SQLC_pdf}
We need to derive the pdf for two cases:

\textbf{Case 1:} Assume that $\rho_x$ is small enough so the limitation to
$\pm \kappa$ is significant. Further let $y_2=f_2(x_2)$, and
$u(\cdot)$ be the Heaviside function. Then the pdf of $y_2$ is
\begin{equation}\label{e:ch2_pdf}
\begin{split}
p_{y_2}(y_2)&=\frac{1}{\sqrt{2\pi}\alpha\sigma_x} e^{-\frac{y_2^2}{2\alpha^2\sigma_x^2}}u(y_2+\alpha\kappa)u(-y_2+\alpha\kappa)+p_o \big(\delta(y_2-\alpha\kappa)+\delta(y_2+\alpha\kappa)\big),
\end{split}
\end{equation}
where
\begin{equation}\label{e:po}
p_o=Pr\{x_2\geq\kappa\}=\int_\kappa^\infty p_x(x_2)\mbox{d}x_2.
\end{equation}
The product $u(y_2+\alpha\kappa)u(-y_2+\alpha\kappa)$ takes the
clipping to $\pm\kappa$ into account. Since samples with
amplitude values outside $[-\kappa,\kappa]$ are represented by $\pm\kappa$ after clipping, one will get an accumulation of
probability mass at $\pm\kappa$, hence the term $p_o
\big(\delta(y_2\pm\alpha\kappa)\big)$. Finally, a
relation from~\cite[pp.131]{papoulis02} was applied to determine the pdf of a
random variable scaled by $\alpha$.

When $f_1$ and $f_2$ are summed, the resulting pdf in (\ref{e:ch2_pdf}), become centered at the transmitted centroid $q_i$ from
encoder 1 (see Fig.~\ref{fig:sqlc_config}), and so the mean of the pdf, given $q_i$, is equal to $\mu$ in~(\ref{e:ch_pdf_mean}).
Let $z_2$ denote the
received signal given $q_i$. Assuming
that the noise is additive (no threshold effects occurring for source 2), the resulting distribution after addition of noise is given by
convolution~\cite[181-182]{papoulis02}
\begin{equation}\label{e:pdf_z}
\begin{split}
&p_{z_2}(z_2)_\kappa =p_{y_2}\ast p_n = \\
&\frac{1}{2\pi\alpha\sigma_x\sigma_n}\int_{-\alpha\kappa}^{\alpha\kappa} e^{-\frac{\alpha^2\sigma_x^2 (z_2-\mu-y)^2+\sigma_n^2 y^2}{2\alpha^2\sigma_x^2\sigma_n^2}}\mbox{d}y+p_o\big(p_n(z_2-\mu-\alpha\kappa)+p_n(z_2-\mu+\alpha\kappa)\big).
\end{split}
\end{equation}

\textbf{Case 2:} The pdf in~(\ref{e:pdf_z}) must be modified when $\rho_x$ gets close to one, since clipping becomes negligible. We can now assume $y_2=f_2(x_2)\approx\alpha x_2$. When $\rho_x >> 0$, each source segment will no longer be equivalent but contain different (but intersecting) \emph{ranges} of $y_2$ (seen from the $\rho_x=0.9$ case in Fig.~\ref{fig:sqlc_config}).
This implies that the channel segments shown in
Fig.~\ref{fig:sqlc_config} are no longer exact copies of each
other but describe somewhat different ranges of $y_2$.
Given that $q_i$ was transmitted, one can show that
\begin{equation}\label{e:pdf_gen_rho_pre}
p(y_2|q_i)=\frac{1}{\sigma_x \alpha \sqrt{2 \pi (1-\rho_x^2)}}e^{-\frac{(y_2-\mu)^2}{2 \alpha^2 \sigma_x^2 (1-\rho_x^2)}},
\end{equation}
by using the expression for $p(x_2|x_1)$~\cite[p.223]{papoulis02}, the scaling of a random variable~\cite[pp. 131]{papoulis02} and inserting $\mu$ from~(\ref{e:ch_pdf_mean}). After addition of noise, the pdf is given by the convolution $p_{z_2 i}(z_2)_\gamma = p(y_2|q_i)\ast
p_n(n)$ (assuming that threshold effects are absent)
\begin{equation}\label{e:pdf_zr}
\begin{split}
p_{z_2 i}(z_2)_\gamma &= \frac{e^{-\frac{1}{2}\frac{(\mu-z_2)^2}{(\alpha^2\sigma_x^2(\rho_x^2-1)+\sigma_n^2)}}}{\sqrt{2\pi (\alpha^2\sigma_x^2(1-\rho_x^2)+\sigma_n^2)}} .
\end{split}
\end{equation}

The validity of~(\ref{e:pdf_z}) and~(\ref{e:pdf_zr}) must be determined. Fig.~\ref{fig:SQLC_corr_concept} provides a geometrical picture for the following discussion. Let $l_1$ denote the length of the portion of the $x_2$ axis that contains the \emph{significant probability mass}\footnote{By ``significant probability mass'' we mean that all events except those with very low probability are included.} given $x_1$ (or $q_i$).  $l_1=2\sqrt{\vartheta}\|\mathbf{e}_2\|=2 b\sqrt{\vartheta\lambda_2}$, where $\|\mathbf{e}_2\|=b\sqrt{\lambda_2}$ denote the length of the minor axis of the ellipse depicted in Fig.~\ref{fig:SQLC_corr_concept} (the source space). $b$ ($\approx 4$) is a parameter determining the width of the ellipse shown, and should be chosen so that the significant probability mass is within this ellipse. $\vartheta=(l_1/(2\|\mathbf{e}_2\|))^2$ and depends on $\rho_x$: When $\rho_x=0$ then $\vartheta = 1$ since the source space is rotationally invariant (a circle). When $\rho_x > \approx 0.7$ then $\vartheta\approx 1/\cos^2 (\psi)=1/\cos^2 (\pi/4)=2$. That is, $\vartheta\in[1,2]$. The pdf in~(\ref{e:pdf_z}) is therefore valid when $l_1 > 2\kappa$
while~(\ref{e:pdf_zr}) is valid when $l_1 \leq 2\kappa$. The total channel output pdf is then given by
\begin{equation}
p_z(z)=\sum_{i= -\infty}^\infty Pr(q_i){p_{z_2 i}(z)}
\end{equation}
where $p_{z_2 i}(z)$ is either~(\ref{e:pdf_z}) or~(\ref{e:pdf_zr}) depending on whether $l_1 > 2\kappa$ or not. To calculate the distortion, one
only need to consider~(\ref{e:pdf_z}) or~(\ref{e:pdf_zr}) centered at the origin, as explained later.
\subsubsection{Distortion and power calculation for source 2}\label{ssec:calc_dist2}
The distortion for source 2 consists of three contributions:
clipping distortion, channel distortion and anomalous distortion.

\textbf{Clipping distortion}: Distortion from clipping, $\bar{\varepsilon}_{\kappa}^2$, results
whenever $|x_2|>\kappa$. That is, an event with probability
$Pr\{|x_2|>\kappa\}$ and resulting error $(x_2-\kappa)^2$.
Therefore 
\begin{equation}\label{e:cutoff_noise}
\bar{\varepsilon}_{\kappa}^2=2\int_\kappa^\infty (x_2-\kappa)^2 p_x(x_2)\mbox{d}x_2.
\end{equation}

\textbf{Channel distortion}:
Consider the effect of channel noise in the absence of threshold
effects and let $\tilde{x}_2=\ell_\pm(x_2)$ denote the clipped source. Then
\begin{equation}\label{e:ch2_noise}
\begin{split}
\bar{\varepsilon}_{Ch2}^2 &= E\{(\tilde{x}_2-(\alpha\tilde{x}_2+n)\beta)^2\}\approx\sigma_x^2(1-\alpha\beta)^2+\beta^2\sigma_n^2,
\end{split}
\end{equation}
where the last approximation comes from assuming that
$E\{\tilde{x}_2^2\}=\sigma_x^2$.

\textbf{Anomalous distortion}: When the wrong centroid from encoder 1 is
detected, large decoding errors result for source 2. In the worst case,
large positive and negative values are interchanged. This can be
seen from Fig.~\ref{fig:sqlc_config}. The magnitude of the error
depends on whether $l_1 > 2\kappa$ or not and is most severe when $\rho_x = 0$.

Consider first that $l_1 > 2\kappa$: The probability for the event $Pr\{x_2+n \geq \Delta/2\}$, which result in anomalies for $x_2$,
is equal for each channel segment, given $q_i$, and can therefore be calculated by
assuming that $q_i=0$
\begin{equation}\label{e:th1}
\begin{split}
p_{th1}&=Pr\bigg\{y_2+n\geq \frac{\Delta}{2}(1+\alpha\rho_x)\bigg\}=2 \int_{\frac{\Delta}{2}(1+\alpha\rho_x)}^\infty p_{z_2}(z_2)_{\kappa|q_i=0}\mbox{d}z_2,
\end{split}
\end{equation}
where $p_{z_2}(z_2)_\kappa$ is given in~(\ref{e:pdf_z}). The error is bounded by $(2\kappa)^2$, since
$\kappa$ is detected as $-\kappa$ when neighboring segments in the
channel space first start to intersect.

Now consider the case when $l_1 \leq 2\kappa$. ~(\ref{e:pdf_zr}) is
just shifted according to which $q_i$ was transmitted from encoder 1,
implying that the probability for anomalies is the same for each
channel segment given $q_i$. I.e. the relevant
probability can again be calculated by assuming that $q_i=0$:
\begin{equation}\label{e:th2}
p_{th2}=2 \int_{\frac{\Delta}{2}(1+\alpha\rho_x)}^\infty p_{zi}(z_2)_{\gamma |q_i=0}\mbox{d}z_2,
\end{equation}
where $p_{zi}(z_2)_\gamma$ is given in~(\ref{e:pdf_zr}). We now need to determine the magnitude of the anomalous errors $\gamma$. Since $\gamma$ is approximately the same in
magnitude no matter which channel segment we ``jump from'' (see Fig.~\ref{fig:SQLC_corr_concept}), we choose to
calculate $\gamma$ by considering jumps between the segments closest to the origin in the source (and channel) space. Then the parallelogram shown to the right in Fig.~\ref{fig:SQLC_corr_concept} can be used to approximately determine $\gamma$.  Since $\Psi=\pi/4$,  the parallelogram consists of a square
and two right triangles with both edges equal to $\Delta$, implying that
\begin{equation}
\gamma\approx l_1-\Delta=2 b\sigma_x \sqrt{\vartheta(1-\rho_x)}-\Delta.
\end{equation}
where $\vartheta\approx2$ since $\rho_x$ is close to one (see Section~\ref{ssec:derive_SQLC_pdf}).

Since $p_{th1}$ and $p_{th2}$ are the same for each segment, the anomalous distortion becomes
\begin{equation}
\bar{\varepsilon}_{an}^2\leq
\begin{cases}
4 p_{th1} \kappa^2, & l_1 > 2\kappa, \vspace{0.3cm}\\
p_{th2}\gamma^2, & l_1 \leq 2\kappa.
\end{cases}
\end{equation}

\textbf{Power}: The transmitted power from encoder 2 is
\begin{equation}\label{e:power_2}
P_2 = \int_{-\infty}^{\infty} y_2^2 p_{y_2}(y_2)\mbox{d}y_2 = \int_{-\alpha\kappa}^{\alpha\kappa}y_2^2 p_{y_2}(y_2)\mbox{d}y_2+2 p_o \alpha^2\kappa^2,
\end{equation}
where $p_o$ is given in~(\ref{e:po}) and the term $2 p_o
\alpha^2\kappa^2$ accounts for the accumulation of probability mass
at $\pm\kappa$ due to clipping. When $\kappa$ is large ($\rho_x$ close to 1), then $P_2\approx \alpha^2\sigma_x^2$.

\subsubsection{Distortion and power calculation for source 1}\label{ssec:calc_dist1}
The distortion for source 1 consists of two contributions:
Quantization distortion and channel distortion.

\textbf{Quantization distortion}: We assume a large enough number of
quantization levels so that the quantization distortion consists
of granular noise only, and we get
\begin{equation}\label{e:q_noise}
\begin{split}
&\bar{\varepsilon}_{q}^2
= 2\sum_{i=1}^{\infty} \int_{(i-1)\Delta}^{i\Delta} \big(x_1-q_i\big)^2p_x(x_1)\mbox{d}x_1 = 2\sum_{i=1}^{\infty} \int_{(i-1)\Delta}^{i\Delta} \bigg(x_1-(i-1)\Delta-\frac{\Delta}{2}\bigg)^2 p_x(x_1)\mbox{d}x_1.
\end{split}
\end{equation}

\textbf{Channel distortion}: Distortion from channel noise, $\bar{\varepsilon}_{C1}^2$, only occur when $x_2+n \geq \Delta/2$.
Since we are only interested in determining the optimal design parameters for
the SQLC, we simplify the analysis by considering jumps to
the nearest neighboring centroids only. The probability for this
event is the same as for the anomalous
distortion calculated in~(\ref{e:th1}) and~(\ref{e:th2}).
The error we get when two neighboring centroids are
interchanged is $\Delta^2$, thus
\begin{equation}
\bar{\varepsilon}_{Ch1}^2= \Delta^2
\begin{cases}
p_{th1}, & l_1 > 2\kappa, \vspace{0.3cm}\\
p_{th2}, & l_1 \leq 2\kappa.
\end{cases}
\end{equation}

\textbf{Power:} The average transmit power for $x_1$ is
\begin{equation}\label{e:power_1}
P_1=2\sum_{i=1}^{\infty} p_i\cdot q_i^2=2\sum_{i=1}^{\infty} p_i \bigg((i-1)\Delta+\frac{\Delta}{2}\bigg)^2, \hspace{0.5cm} p_i=\int_{(i-1)\Delta}^{i\Delta} p_x(x_1)\mbox{d}x_1.
\end{equation}

\subsubsection{Optimization}\label{ssec:optimize_sqlc}
\begin{figure}[h!]
    \begin{center}
    \subfigure[]{
            \includegraphics[width=0.6\columnwidth]{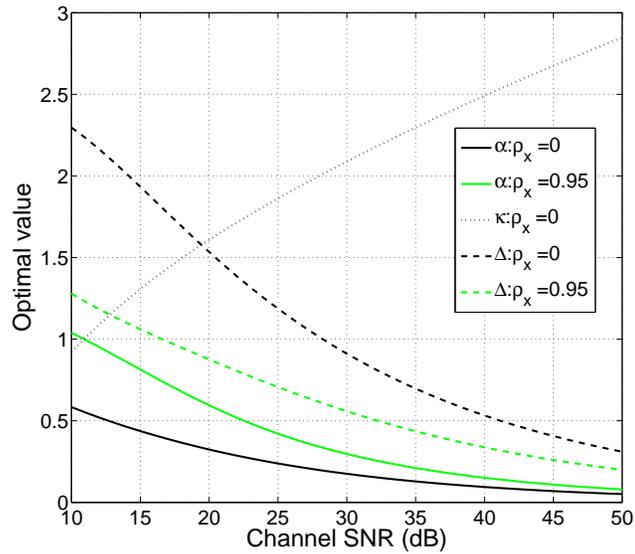}
        \label{fig:opt_alp_delt_kap}}
        \hfil
        \subfigure[]{
            \includegraphics[width=0.6\columnwidth]{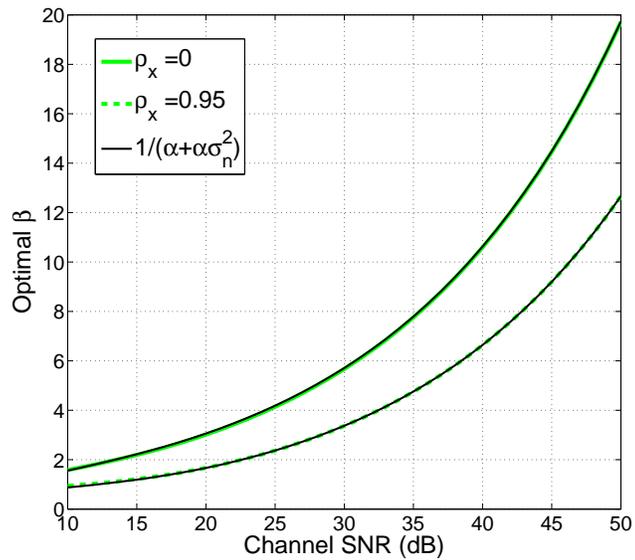}
        \label{fig:opt_beta}}
    \end{center}
    \caption{Optimal parameters for the SQLC system when $\rho_x = 0$ and $0.95$ ($P=1$ and $\sigma_x=1$). The results shown assume a scaling of $x_1$ by $\sqrt{2 P -P_2}/\sigma_x$ prior to quantization.
    ~\ref{fig:opt_alp_delt_kap} $\alpha$, $\Delta$ and $\kappa$. ~\ref{fig:opt_beta} $\beta$.}\label{fig:opt_par}
\end{figure}
Most of the terms derived in section~\ref{ssec:calc_dist2} and~\ref{ssec:calc_dist1} can be given closed form expressions by applying the \emph{error function}, except for terms including the integral
in~(\ref{e:th1}) which must be solved numerically. Numerical optimization is therefore necessary in order to determine the optimal parameters. Instead of solving the constrained problem in~(\ref{e:opt_prob_sqlc}), we choose to scale $x_1$ by $\xi = \sqrt{2P-P_2(\alpha,\kappa)}/\sigma_x$ prior to quantization so we get an unconstrained problem instead. Note that with this scaling we must modify the factor $\mu$ with $q_i(1+\alpha\rho_x/\xi)$ in~(\ref{e:ch_pdf_mean}) to decode correctly.
The average distortion now becomes $D=(D_1+D_2)/2$ where
\begin{align}\label{e:opt_prob_sqlc}
&D_1(\alpha,\Delta,\kappa)=\frac{\varepsilon_{q}^2(\Delta)+\varepsilon_{Ch1}^2(\alpha,\Delta,\kappa)}{2P - P_2(\alpha,\kappa)}\nonumber\\
&D_2(\alpha,\beta,\Delta,\kappa)=\varepsilon_{\kappa}^2(\kappa)+\varepsilon_{Ch2}^2(\alpha,\beta)+\varepsilon_{an}^2(\alpha,\Delta,\kappa).
\end{align}
The optimized parameters as a function of the channel SNR is
depicted in Fig.~\ref{fig:opt_par} for $\rho_x = 0$ and $\rho_x=0.95$.
Notice that when $\rho_x=0.95$, we get a smaller $\Delta$ and a
larger $\alpha$ for a given channel SNR compared to the $\rho_x=0$
case, implying improved fidelity (SDR) when $\rho_x$ increases.
$\kappa$ is not plotted in Fig.~\ref{fig:opt_alp_delt_kap} for
$\rho_x=0.95$, since it becomes irrelevant.
Notice also that $\beta$ fits quite well to the function $1/(\alpha+\alpha\sigma_n^2)$ when SNR $>10$ dB. It is therefore simple to find decoder 2 once encoder 2 is known. The curves in
Fig.~\ref{fig:opt_alp_delt_kap} can be given mathematical expressions
using nonlinear curve fitting to e.g. exponential
functions. With channel state information available one can adapt the encoders to varying channel conditions
simply by using these functions.
\subsection{Comparison and extension of NQ and SQLC}\label{ssec:comp_nq_sqlc}
From Section~\ref{ssec:nest_q} and~\ref{ssec:SQLC} one can observe that the NQ and
the SQLC have similarities as well as differences. The main difference lies in how the encoding is performed for $x_2$: The NQ is discrete and use a sawtooth-like mapping $\tilde{i}_2(x_2)$ to limit the output of encoder 2
to a fixed bounded interval, whereas the SQLC is continuous and either clips the amplitudes of $x_2$ (at low $\rho_x$) or simply rely on the fact that $p_\mathbf{x}(x_1,x_2)$
limits the amplitudes of $x_2$ for a given $q_i$  when $\rho_x$ is close to one.
However, if the different parameters of the two schemes ($c$, $\Delta$, $\alpha$ and $\kappa$) are chosen optimally (so that anomalous errors are avoided), the geometrical configuration of the NQ after correct reconstruction at the decoder is quite similar to a quantized version of the
SQLC, at least at high SNR. One may therefore expect the optimal behavior of the NQ and SQLC to be quite similar, and that the NQ have
a certain loss compared to the SQLC due to an additional quantization distortion term.

Although the SQLC has better performance than NQ, the NQ will be advantageous when a digital system must be constructed. Further, by letting $\Delta\rightarrow 0$ for the NQ in encoder 2 we get a hybrid discrete-analog scheme where $\tilde{i}_2$ is replaced by a continuous sawtooth-like function, as that in~\cite{Yao_Skoglund_sawtooth}.
The resulting scheme should at least improve over the NQ. The question is if this modified NQ approach can improve upon the SQLC. Further research is needed.

In the boundary case $\rho_x\rightarrow 1$,
we can let $\Delta\rightarrow 0$ and $c\rightarrow 1$ for the NQ and
$\Delta\rightarrow 0$, $\kappa\rightarrow\infty$ and $\alpha=1$ for
the SQLC. This means that both schemes are reduced to a
distributed linear mapping, i.e. uncoded transmission, which is the
optimal communication strategy when $\rho_x=1$~\cite[Corollary IV.3]{Lapidoth10}\footnote{For the NQ to reach the bound we must let $P_1=2P$ since $\tilde{i}_2=0$ when $\rho_x=1$. This is achieved by setting $a=\sqrt{2P}$.}.

The SQLC and NQ can be optimized using the same procedure provided in this paper for any unimodal source distribution.  The performance will depend on the tails of the distribution. That is, heavier tails than the Gaussian should result in worse performance, whereas smaller tails should result in improved performance. Consider a uniform v.s. a Laplacian distribution for the SQLC: In the uniform case the SDR will increase for
a given SNR compared to the Gaussian since the pdf is narrower.
One can also avoid distortion from clipping since the uniform
distribution has compact support. In the Laplacian case, one must
expect a loss in SDR compared to the Gaussian case, since large
amplitude values have higher probability. This will result in a
larger clipping distortion and/or a lower resolution for the
quantizer of encoder 1.

Both schemes can rather easily be extended to he multivariate case, i.e. when $M$ sources are assumed.
Also, increasing the codelength beyond zero delay is rather straight forward by applying vector quantizers and linear coders of dimension $N$.
%
\subsection{High SNR analysis}
The upper bound~(\ref{e:OPTA_equalP}) can not be achieved with either the NQ
or the SQLC even when SNR $\rightarrow\infty$ (except when $\rho=1$). One reason is that
both schemes apply scalar quantization which is sub-optimal. Both schemes do, however, exhibit a constant gap to the bound as the SNR$\rightarrow\infty$. We will quantify this gap for the SQLC in the following. The reason why the NQ displays similar behavior will be briefly explained afterwards.

The analysis here is \emph{approximate} since, contrary to the case of infinite code length, there is a significant variance around the mean length of any stochastic vector due to short code length~\cite[p. 324]{wozandj65} (for a normalized i.i.d. Gaussian random vector $\bar{\mathbf{x}}$ of dimension $N$ we have $\text{Var}\{\|\bar{\mathbf{x}}\|\}=2\sigma_x^4/N$). Further, to find closed form expressions that we can analyze further, we do not take all distortion terms into account, only the ones dominant under close to optimal conditions at high SNR: We can (nearly) avoid anomalous errors by assuming a distance $\Delta > \sqrt{\alpha^2 l_1^2 + \delta_n^2}$ between each centroid in Fig.~\ref{fig:sqlc_config} (the distance is actually $\Delta(1+\alpha \rho_x) > \sqrt{\alpha^2 l_1^2 + \delta_n^2}$, but since $\Delta(1+\alpha \rho_x)>\Delta$, our assumption is still valid. The extra term is also of little significance when the SNR is large since $\alpha$ is very small). $l_1=2 b \sqrt{\vartheta (1-\rho_x)}$, as shown in Section~\ref{ssec:derive_SQLC_pdf}, and $\delta_n= 2 b_n \sigma_n$, where $b_n$ ($\approx 4$) is a constant that must be chosen so that the significant probability mass of the noise $n$ is within $2 b_n \sigma_n$. Assuming that clipping gets negligible when SNR grows large and noting that $\delta_n \rightarrow 0$ as SNR$\rightarrow \infty$, the distortion can be approximated by
\begin{equation}\label{e:dist_high_snr}
D_1 \approx \frac{\Delta^2}{12\big(2P-P_2\big)}=\frac{\sigma_x^2 b^2 \vartheta \alpha^2 (1-\rho_x)}{3\big(2P - \alpha^2 \sigma_x^2\big)}, \hspace{1cm} D_2\approx\frac{\sigma_n^2}{\alpha^2}.
\end{equation}
We have scaled $D_1$ by $2 P - \alpha^2 \sigma_x^2$ to satisfy the average power constraint, and applied the high rate approximation for a scalar quantizer.
For $D_2$ we have used the high SNR approximation (ML decoding) of~(\ref{e:ch2_noise}). By solving ${\partial \big[D_1+D_2\big]}/{\partial \alpha} =0$
with respect to $\alpha^2$, we get
\begin{equation}\label{e:opt_alph_high_SNR}
\alpha^2 =\frac{2 P \big(b\sqrt{6 \vartheta \text{SNR} (1-\rho_x)}-3\sigma_x\big)}{\sigma_x \big(2 b^2 \vartheta \text{SNR} (1-\rho_x) -3\sigma_x^2\big)}\approx\frac{P \sqrt{6 \vartheta \text{SNR} (1-\rho_x)}}{\sigma_x  b \vartheta \text{SNR} (1-\rho_x) }, \hspace{1cm}\rho_x \neq 1,
\end{equation}
where SNR$=P/\sigma_n^2$, and the last approximation results from removing constant terms at high SNR. We assume that $\sigma_x=1$ in the following for simplicity.
By inserting~(\ref{e:opt_alph_high_SNR}) in~(\ref{e:dist_high_snr}) and assuming that the SNR is very large, one can show that
\begin{equation}
\begin{split}
&\frac{1}{D_1} =\frac{3\big(2P/\alpha^2 -1\big)}{b^2 \vartheta (1-\rho_x)} \approx \frac{6 b \vartheta \text{SNR} (1-\rho_x)}{b^2 \vartheta (1-\rho_x)\sqrt{6 \vartheta \text{SNR}(1-\rho_x)}}=\frac{\sqrt{3}}{b\sqrt{\vartheta}} \sqrt{\frac{2\hspace{0.5mm} \text{SNR}}{1-\rho_x}},  \hspace{1cm}\rho_x \neq 1,\\
&\frac{1}{D_2} =\frac{\alpha^2}{\sigma_n^2} \approx\frac{\text{SNR}\sqrt{6 \vartheta \text{SNR} (1-\rho_x)} }{b \vartheta \text{SNR} (1-\rho_x)}= \frac{\sqrt{3}}{b \sqrt{\vartheta}} \sqrt{\frac{2\hspace{0.5mm} \text{SNR}}{1-\rho_x}},  \hspace{1cm}\rho_x \neq 1.
\end{split}
\end{equation}
Since the upper bound can be approximated by $\text{SDR}_{UB}\approx \sqrt{2\text{SNR}/(1-\rho_x)}$,  $\rho_x \neq 1$,  when SNR is large~\cite{Lapidoth10},
the loss from the bound is approximately quantified by $\text{SDR}_{loss}\approx \sqrt{3}/(b\sqrt{\vartheta})$.
By inserting $b\approx 4$ and $\vartheta\approx 1$ when $\rho_x$ is close to zero and $\vartheta \approx 2$ when $\rho_x$ is close to one, we find that the distance to the upper bound is around 3.5 dB when $\rho_x=0$ and around 5 dB when $\rho_x=0.95$.
This estimate is somewhat pessimistic when $\rho_x$ is close to zero (by $\approx 0.5$ dB) but relatively accurate when $\rho_x$ is close to one (will be evident from the simulations in Section~\ref{sec:perf_and_sim}).
 The estimate anyway indicates that there is a constant gap to the bound when the channel SNR grows large.

Since the SQLC has a constant gap to the bound and the NQ is similar to a quantized version of the SQLC under optimal conditions (see Section~\ref{ssec:comp_nq_sqlc}) it naturally follows that also the NQ exhibit a constant gap to the bound.
The gap will be somewhat larger since the NQ has an additional quantization distortion term. A similar method as that derived in~\cite[83-102]{HeklandThesis}
may be applied to quantify this gap, but is rather involved and will not be included here.
\section{Simulations}\label{sec:perf_and_sim}
We compare the NQ and SQLC to the
collaborative S-K mappings, the performance (SDR) upper bound
from Section~\ref{ssec:prob_def_b}, uncoded transmission and the SSCC bound (derived in~\cite{Lapidoth10}).

We consider both an average- and equal transmit power constraint, where $P=1$ in both cases.
We will further assume that $\sigma_x=1$ and use the optimal parameters resulting from the optimization problem in~(\ref{e:opt_problem_nq}) for the NQ and the optimal parameters from Section~\ref{ssec:optimize_sqlc}
for the SQLC. Note that for Uncoded transmission, S-K mappings and the SSCC bound, $P_1=P_2=P$ leads to the best possible performance under an average transmit power constraint.
\subsection{Average transmit power constraint}
\begin{figure}[h!]
    \begin{center}
        \subfigure[]{
            \includegraphics[width=0.6\columnwidth]{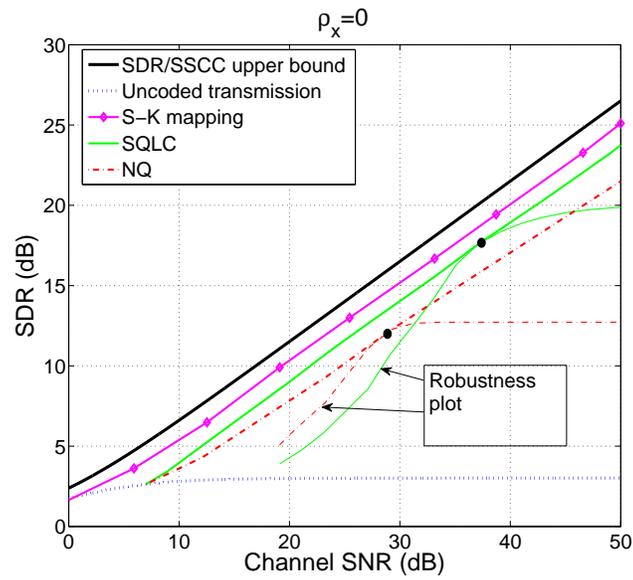}
        \label{fig:perf_r0}}
        \hfil
        \subfigure[]{
            \includegraphics[width=0.6\columnwidth]{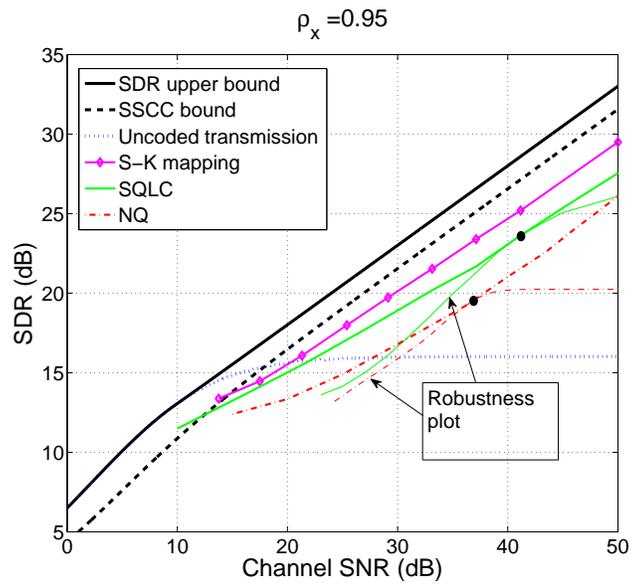}
        \label{fig:perf_r095}}
    \end{center}
    \caption{Comparison of relevant schemes with an average power constraint. The dots shows the design SNR for the robustness plots.~\ref{fig:perf_r0} $\rho_x=0$.~\ref{fig:perf_r095} $\rho_x=0.95$.}\label{fig:tot_pow_results}
\end{figure}

Simulation results are shown in Fig.~\ref{fig:tot_pow_results} for
$\rho_x=0$ and $\rho_x=0.95$.

When $\rho_x=0$ (Fig.~\ref{fig:perf_r0}) the SQLC is about
2-3 dB away from the performance (SDR) upper bound and the NQ is inferior to the SQLC by
about 0-2 dB.
Both schemes are significantly better than uncoded transmission, but inferior to the S-K mapping, which is
only 1-1.5 dB away from the upper bound. One reason why the SQLC backs off from the S-K mapping is that the
S-K mapping is continuous and therefore completely avoid threshold effects~\cite[pp.30-32]{FloorThesis}. This leads to a better utilization of the channel space, since the SQLC must leave an empty interval between each channel segment in order to avoid
threshold effects.
\begin{figure}[h!]
    \begin{center}
        \subfigure[]{
            \includegraphics[width=0.6\columnwidth]{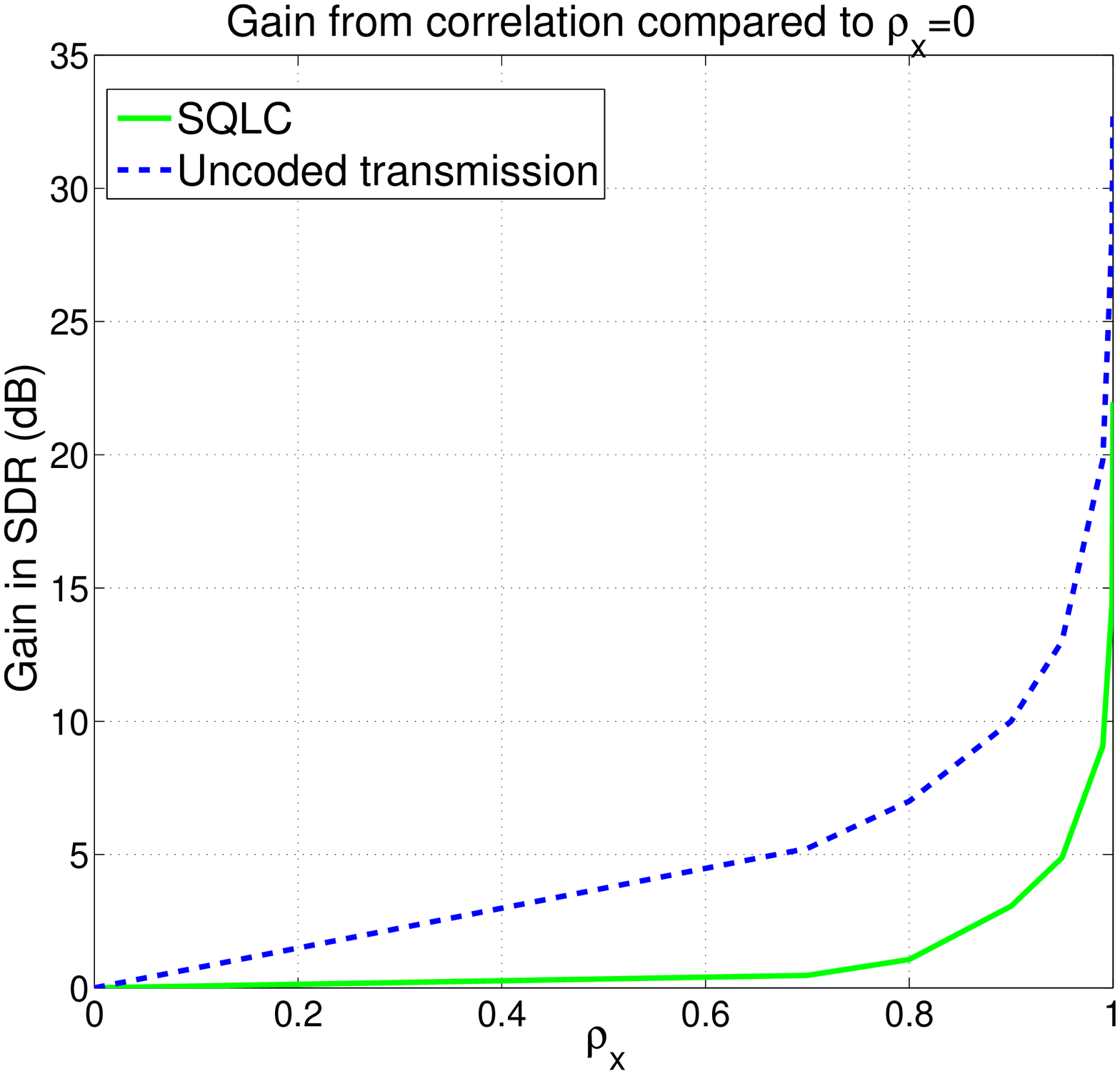}
        \label{fig:gain_rho_sqlc}}
        \hfil
        \subfigure[]{
            \includegraphics[width=0.6\columnwidth]{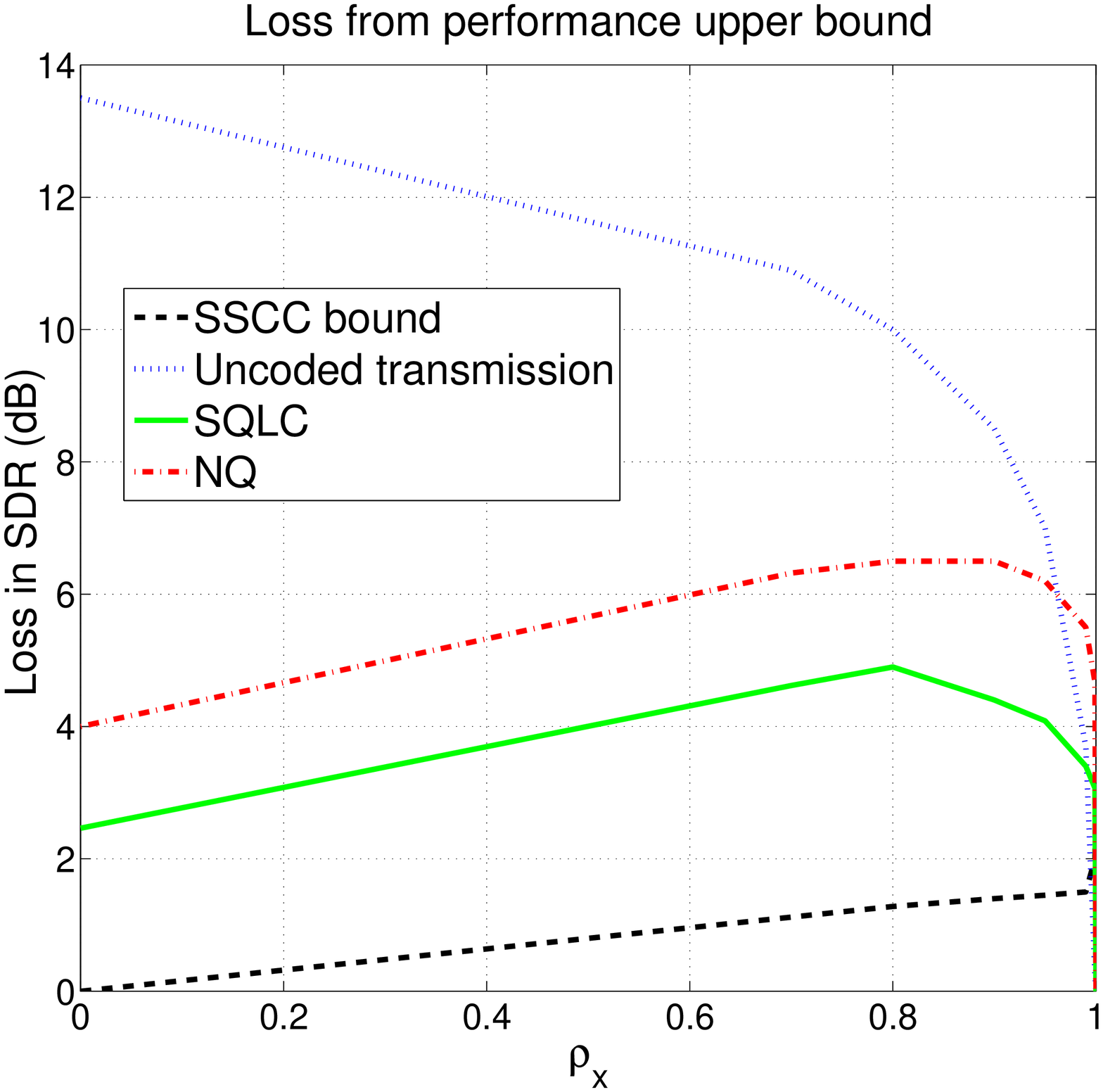}
        \label{fig:loss_opta_rho}}
    \end{center}
    \caption{How correlation affects performance.~\ref{fig:gain_rho_sqlc}
    Gain from correlation for the SQLC and uncoded transmission.~\ref{fig:loss_opta_rho}
    Loss from OPTA as a function of $\rho_x$ for 30 dB channel SNR.}\label{fig:Correlation_influence}
\end{figure}
When $\rho_x=0.95$ (Fig.~\ref{fig:perf_r095}) the performance of all
schemes improve in SDR. The gap to the upper bound in terms of SDR,
however, becomes larger for the S-K mapping ($\approx 2-3$dB), SQLC
($\approx  2.5-5$dB) and NQ ($\approx 3-7$dB). The contrary is true for uncoded
transmission.
Considering that the mappings are delay free, the
performance is still quite good. Interestingly, both the SQLC and NQ improve with
increasing $\rho_x$ without modification of the basic encoding and decoding
structure, i.e only the parameters $\Delta,\kappa,\alpha,\beta,c$
need to be changed.

Robustness plots are also displayed in Fig.~\ref{fig:tot_pow_results}. The black dots mark the designed SNR: 29 dB for NQ and 37 dB for SQLC when $\rho_x=0$, and 37 dB for NQ and 41 dB for SQLC when $\rho_x=0.95$. Both schemes improve and degrade
gracefully under a channel SNR mismatch.

The gain from increasing correlation as a
function of $\rho_x$ is shown in Fig.~\ref{fig:gain_rho_sqlc} for
the SQLC and uncoded transmission at 30 dB channel SNR.
Note that the gain for the SQLC is not significant before $\rho_x > \approx
0.7$, whereas the gain gets large when $\rho_x\rightarrow 1$. Uncoded transmission shows an even greater gain, which is natural since it goes from being highly sub-optimal when $\rho_x=0$  to achieve the bound for all SNR when $\rho_x=1$.
The gap to the performance upper bound as a function of $\rho_x$ is
plotted for NQ, SQLC, uncoded transmission and SSCC bound in
Fig.~\ref{fig:loss_opta_rho}, for 30dB channel SNR. Note that the distance to the upper bound is largest for both the NQ and SQLC when $\rho_x$ is around $0.8$ and that SQLC, NQ and uncoded transmission all reach the upper bound in the limit
$\rho_x\rightarrow 1$, whereas the SSCC bound does not.

\subsection{Equal transmit power constraint}\label{ssec:sim_equal_power}
Under an equal transmit power constraint, there is a loss in
performance for both  SQLC and NQ since they have asymmetric encoders. To simulate equal transmit power we use \emph{time sharing}:
Each source is encoded by $f_1(\cdot)$ half the time and
$f_2(\cdot)$ the other half. Averaged over a large number of samples, then $P_1=P_2=P$ is achieved.
\begin{figure}[h!]
    \begin{center}
              \includegraphics[width=0.6\columnwidth]{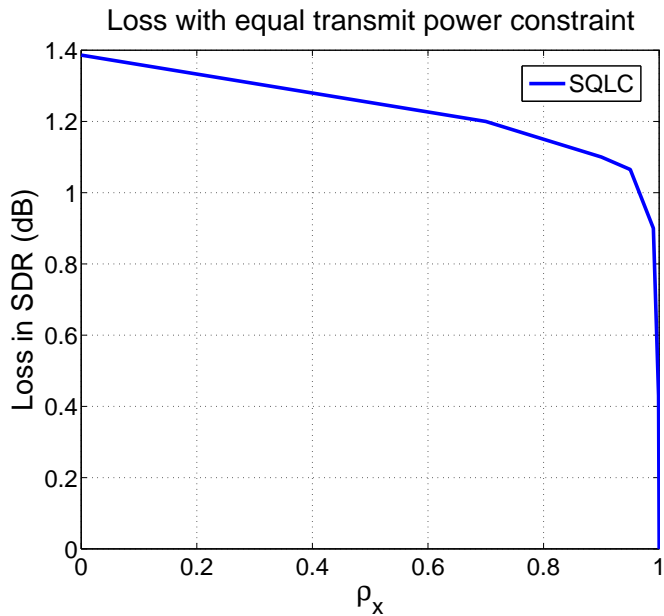}
    \end{center}
    \caption{How an equal transmit power constraint $P_1=P_2$ affects
    performance for the SQLC.
    The loss is compared to the average power constraint case with optimal $P_1$ and $P_2$.
    The loss is shown as a function of $\rho_x$ when the channel SNR$=30$ dB.}\label{fig:loss_tx_constraint_all_rho}
\end{figure}
Fig.~\ref{fig:loss_tx_constraint_all_rho} shows the loss in
performance for the SQLC with an equal power constraint compared
to an average power constraint as a function of $\rho_x$  at 30 dB SNR (a similar effect is observed for the NQ).
Notice that the loss becomes less as $\rho_x$ increases since the difference in power $P_1-P_2$ becomes smaller. When $\rho_x = 1$, no loss is observed since the two
encoders are equal.
\section{Summary and future Research}\label{sec:summary}
In this paper, distributed delay free joint source-to-channel mappings for a bivariate Gaussian communicated on a Gaussian multiple
access channel (GMAC) were proposed. We optimized a discrete mapping
based on nested quantization (NQ) and a hybrid discrete-analog
scheme SQLC which incorporates a piecewise continuous mapping.
Both schemes are well performing and improve on uncoded transmission for most SNR values outside the domain where uncoded transmission is optimal, thereby closing some of the gap to the performance upper bound.
Since the NQ and SQLC have asymmetric encoders, a certain loss is observed when an equal transmit power constraint is imposed. Both NQ and the SQLC improve
with increasing $\rho_x$ without changing the basic structure of the
encoders and decoders, and achieves the performance upper bound when $\rho_x\rightarrow 1$.

A collaborative scheme using Shannon-Kotel'nikov mappings was also discussed to provide an indication on where the bound for zero-delay coding may lie. Naturally, a back-off from this scheme was observed for both the SQLC and NQ. The main reason being that better mappings can be constructed when the encoders cooperate, i.e. mappings that make it possible to avoid so-called threshold effects. The SQLC is not necessarily the optimal distributed scheme either which may result in an additional loss factor.

In our current research, we consider multiple sources as well as
different source statistics and different attenuation for each
sub-channel of the GMAC. One approach may be a generalization of the NQ and SQLC scheme. It would also be beneficial to determine
if there exists better ways of doing zero delay distributed coding than the NQ and SQLC.

Identifying the optimal performance of a communication system with
a finite dimensionality constraint is crucial in assessing the
performance of the proposed JSCC schemes. It is, however, a very
difficult problem since many of the standard tools in information
theory rely on infinitely long codewords. A generalization of work
such as~\cite{Verdu09} could be an important step towards finding
the performance bounds under such constraints.

\appendix
Using $\hat{x}_m$ and the auxiliary variables $\bar{x}_m$ and $\tilde{x}_m$ from~(\ref{e:aux_var}), $D_m$,  $m \in \{1,2 \}$, can be written
\begin{equation}
\begin{split}
    &D_m = E\{(x_m - \hat{x}_m)^2\}= E\{(x_m - \bar{x}_m + \bar{x}_m  - \tilde{x}_m +
       \tilde{x}_m - \hat{x}_m)^2\}\\
       &={E\{(x_m - \bar{x}_m)^2\}} + {E\{(\bar{x}_m - \tilde{x}_m)^2\}}  +    {E\{(\tilde{x}_m - \hat{x}_m)^2\}}+ 2 {E\{(x_m - \bar{x}_m)(\bar{x}_m  - \tilde{x}_m)\}}\\
        &+ 2 {E\{(x_m - \bar{x}_m)(\tilde{x}_m - \hat{x}_m)\}} + 2 {E\{(\bar{x}_m  - \tilde{x}_m)(\tilde{x}_m -    \hat{x}_m)\}}.
\end{split}
\end{equation}
For the first cross term we get
\begin{equation}
\begin{split}
  &E\{(x_m - \bar{x}_m)(\bar{x}_m  - \tilde{x}_m)\} =\\& \sum_{i_1, i_2} \iint_{x_1, x_2 \in i_1, i_2} p_{\mathbf{x}}(x_1,x_2)(x_m -\bar{x}_m(i_1,i_2)) (\bar{x}_m(i_1,i_2) -
    \tilde{x}_m(i_1,\tilde{i}_2(i_2))) \ud x_1 \ud x_2= \\
    &\sum_{i_1, i_2} (\bar{x}_m(i_1,\tilde{i}_2(i_2)) -
    \tilde{x}_m(i_1,i_2))\iint_{x_1, x_2 \in i_1, i_2} p_{\mathbf{x}}(x_1,x_2) (x_m -
    \bar{x}_m(i_1,i_2))\ud x_1 \ud x_2 = 0,
\end{split}
\end{equation}
since the last integral is $0$ (since the integral is over $x_1,x_2\in i_1,i_2$). For the second cross term we get the
same integral since
\begin{equation}
\begin{split}
    &E\{(x_m - \bar{x}_m)(\tilde{x}_m  - \hat{x}_m)\} = \sum_{i_1, i_2, j_1, j_2} \iint_{x_1, x_2 \in i_1, i_2} p_{\mathbf{x}}(x_1,x_2)  Pr(j_1, j_2|i_1, i_2)\cdots\\&(x_m - \bar{x}_m(i_1,i_2))(\tilde{x}_m(i_1,\tilde{i}_2(i_2)) -\hat{x}_m(j_1,j_2)) \ud x_1 \ud x_2=\\&\sum_{i_1, i_2, j_1, j_2} Pr(j_1, j_2|i_1,i_2)(\tilde{x}_m(i_1,\tilde{i}_2(i_2)) -\hat{x}_m(i_1,i_2))\cdots\\&\iint_{x_1, x_2 \in i_1, i_2} p_{\mathbf{x}}(x_1,x_2) (x_m - \bar{x}_m(i_1,i_2)) \ud x_1 \ud x_2 = 0.
\end{split}
\end{equation}
Finally, studying the third cross term we get
\begin{equation}
\begin{split}
    &E\{(\bar{x}_m - \tilde{x}_m)(\tilde{x}_m  - \hat{x}_m)\} =\sum_{i_1, i_2, j_1, j_2} \iint_{x_1, x_2 \in i_1, i_2} p_{\mathbf{x}}(x_1,x_2)\cdots \\
    &Pr(j_1, j_2|i_1, i_2)(\bar{x}_m(i_1,i_2) - \tilde{x}_m(i_1,\tilde{i}_2(i_2))) (\tilde{x}_m(i_1,\tilde{i}_2(i_2)) - \hat{x}_m(j_1,j_2)) \ud x_1 \ud x_2 =\\
    &\sum_{i_1, i_2, j_1, j_2} Pr(i_1, i_2) Pr(j_1, j_2|i_1,i_2)(\bar{x}_m(i_1,i_2) - \tilde{x}_m(i_1,\tilde{i}_2(i_2)))(\tilde{x}_m(i_1,\tilde{i}_2(i_2)) - \hat{x}_m(j_1,j_2)) =\\
    &\sum_{i_1, i_2} Pr(i_1, i_2) (\bar{x}_m(i_1,i_2) -\tilde{x}_m(i_1,\tilde{i}_2(i_2)))\sum_{j_1, j_2} Pr(j_1, j_2|i_1, i_2)(\tilde{x}_m(i_1,\tilde{i}_2(i_2)) - \hat{x}_m(j_1,j_2)) = 0.
\end{split}
\end{equation}
due to the choice of  $\tilde{x}_m$. What is left is then the three terms in~(\ref{e:Dtot_discrete}) given by the three integrals~(\ref{e:nq_qdist}),~(\ref{e:nq_cdist}) and~(\ref{e:nq_ndist}).
Hence, we conclude that $D_m = \bar{\varepsilon}_{q,m} + \bar{\varepsilon}_{c,m} + \bar{\varepsilon}_{n,m}$.

\bibliographystyle{IEEEtran}
\bibliography{references}
\end{document}